\documentstyle[prd,preprint,aps]{revtex}

\tighten

\begin{document}

\preprint{SFU HEP-131-96}

\draft

\title{Quarkonium spin structure in lattice NRQCD}

\author{Howard D. Trottier}
\address{Department of Physics, Simon Fraser University, 
Burnaby, B.C., Canada V5A 1S6
\footnote{Email address: trottier@sfu.ca.}}

\date{November 1996}

\maketitle

\begin{abstract}
\noindent
Numerical simulations of the quarkonium spin splittings 
are done in the framework of lattice nonrelativistic
quantum chromodynamics (NRQCD). At leading
order in the velocity expansion the spin splittings are
of $O(M_Q v^4)$, where $M_Q$ is the renormalized 
quark mass and $v^2$ is the mean squared quark velocity
($v^2_\psi \approx .3$ and $v^2_\Upsilon \approx .1$).
A systematic analysis is done of all next-to-leading order 
corrections. This includes the addition of $O(M_Q v^6)$ relativistic 
interactions, and the removal of $O(a^2 M_Q v^4)$ 
discretization errors in the leading-order interactions.
Simulations are done for both $S$- and $P$-wave mesons,
with a variety of heavy quark actions and 
over a wide range of lattice spacings. Two prescriptions for 
the tadpole improvement of the action are also studied in detail:
one using the measured value of the average plaquette,
the other using the mean link measured in Landau gauge.
Next-to-leading order interactions result in a very
large reduction in the charmonium splittings, 
down by about 60\% from their values at leading order.
There are further indications
that the velocity expansion may be poorly convergent for
charmonium. Prelimary results show a small correction
to the hyperfine splitting in the Upsilon system.
\end{abstract}

\section{Introduction}

Quarkonium physics has been the subject of renewed 
theoretical interest in recent years. 
The rich phenomenology of the charmonium and Upsilon
families has spurred the development of nonrelativistic 
quantum chromodynamics (NRQCD), an effective field theory
that relies on an expansion of the action in the mean squared
velocity $v^2$ of the heavy quarks
($v^2_\psi \approx .3$ and $v^2_\Upsilon \approx .1$). 

NRQCD has been formulated both on the lattice
\cite{LepThac,N1992} and in the continuum \cite{BBL}.
Lattice simulations of the Upsilon and charmonium systems
have recently been done by the NRQCD collaboration 
\cite{Nbmass,Nups,Npsi,Nups57,Shig96}. Results have
also been reported for heavy-light \cite{UKQCD}
and $b \bar c$ spectra \cite{Nups57}, and some 
unquenched simulations have also been done \cite{Shig96}. 

There are two key theoretical ingredients underlying lattice 
NRQCD calculations. One is an expansion of the effective
action to a sufficient order in the heavy quark velocity,
so as to obtain results of the desired accuracy. The other key 
ingredient is the use of tadpole renormalization \cite{LepMac}
of the operators in the lattice action, which may
then allow for reliable tree-level matching of the lattice
theory with continuum QCD. A related problem is the development
of sufficiently accurate discretizations of the relevant 
operators. 

The quarkonium spin structure is particularly sensitive
to the details of the NRQCD Hamiltonian. The Upsilon \cite{Nups,Nups57}
and charmonium \cite{Npsi} spin splittings were recently analyzed
to leaving order in $v^2$ by the NRQCD collaboration.
At leading order in the velocity expansion
the spin splittings are of $O(M_Q v^4)$,
where $M_Q$ is the renormalized quark mass. 
In the charmonium system however
appreciable next-to-leading-order effects are expected, given
the large mean squared velocity.
Indeed a recent analysis of the charmonium hyperfine splitting
using the relativistic Fermilab action gives a result ($\approx 70$~MeV)
\cite{FNAL} that is significantly smaller than is obtained from
leading-order NRQCD ($\approx 96$~MeV) \cite{Npsi}. 

In this paper a systematic analysis is done of all next-to-leading 
order corrections to the spin splittings. This includes the addition
of $O(M_Q v^6)$ relativistic interactions, as well as the removal
of $O(a^2 M_Q v^4)$ discretization errors that are present in
the leading-order spin-dependent operators considered in 
Refs. \cite{Nups,Npsi,Nups57}. 

Results from simulations with a variety of heavy quark actions
and over a wide range of lattice spacings 
are presented for the charmonium $S$-wave hyperfine splitting.
Some preliminary results are also reported for the charmonium 
$P$-wave fine structure, and for the Upsilon hyperfine splitting.

Furthermore two prescriptions for defining the tadpole
improvement of the action are studied in detail:
one using the measured value of the average plaquette, as 
considered in Refs. \cite{Nups,Npsi}, the other using the mean link
measured in Landau gauge. Landau gauge tadpole improvement
has recently been shown to yield smaller discretization errors
in the gluonic action (as measured by violations of rotational
invariance in the heavy quark potential), compared to
calculations using the average plaquette as input \cite{LepLandau}.
It is also interesting to note that the mean link is maximized 
in Landau gauge, so this prescription provides a lower bound on 
the tadpole renormalization compared to mean link determinations
using other gauge fixings.

An important aspect of these simulations 
is the removal of leading discretization errors in the
gluonic action as well as in the NRQCD action.
Specifically, an $O(a^4)$-accurate gluonic action is used,
together with $O(a^4)$-accurate clover fields and covariant 
derivatives in the heavy quark action. Tadpole improvement of 
both the gluonic and the heavy quark actions has recently 
been shown to give a good description of the spin-averaged
charmonium spectrum even on coarse lattices with spacings
$a$ as large as .4~fm \cite{LepCoarse}.
In this paper the charmonium spin splittings
are computed on lattices with spacings in the range of
about .17~fm to .39~fm.

The next-to-leading order interactions are shown to 
result in a very large reduction in the charmonium hyperfine 
splitting, down by about 60\% from the leading order result 
reported in Ref. \cite{Npsi} on a lattice of comparable spacing
(when the same tadpole improvement scheme is used);
results for the triplet $P$-wave meson masses show that
the fine structure splittings are also reduced by about 60\%.
The next-to-leading order charmonium hyperfine splitting 
is about $(55 \pm 5)$~MeV, compared to the relativistic Fermilab 
action result of about $(70 \pm 3)$~MeV \cite{FNAL}.
While the two calculations have different systematic
errors, this comparsion suggests that further relativistic 
corrections, beyond the next-to-leading order considered here, 
are again very large.

These results indicate that the NRQCD velocity expansion may be 
poorly convergent for charmonium, with the first three terms in the
expansion for the hyperfine splitting apparently oscillating in sign.
Another possibility is that there are large radiative corrections 
to the coefficients of the spin-dependent operators in the effective 
action. Preliminary results for the Upsilon hyperfine splitting at 
next-to-leading order in $v^2$ (and with leading discretization errors 
removed) show little change from the leading-order result reported
in Ref. \cite{Nups57}.

The dependence of the hyperfine splitting on lattice spacing
is also analysed. The splitting shows very little 
dependence on $a$ when Landau gauge tadpole renormalization 
is used. The average plaquette tadpole scheme on
the other hand has large discretization errors, but
the results are not inconsistent with an extrapolation to the
same splitting at zero lattice spacing in the two schemes.
These results provide further support for the 
use of Landau gauge tadpole renormalization \cite{LepLandau}.

\section{Quark and gauge-field actions}

The NRQCD effective action is based on power-counting rules
for the magnitude of heavy quark and gauge field operators
in quarkonium states. The expansion parameters are
the mean squared velocity of the heavy quarks in the
bound state, and the strong coupling constant.
The coefficients of the operators in the
effective action, to a given order in $v^2$, are determined by
matching the predictions of NRQCD with those of full QCD
\cite{LepThac,N1992}.

The heavy quark lattice Hamiltonian is conveniently decomposed 
into the leading covariant kinetic energy operator $H_0$, plus
relativistic and discretization corrections $\delta H$.
Following Refs. \cite{Nups,Npsi}, the quark Green function
is given by
\begin{equation}
   G_{t+1} =
   \left(1\!-\!\frac{aH_0}{2n}\right)^n
   U^\dagger_4
   \left(1\!-\!\frac{aH_0}{2n}\right)^n
   \left(1\!-\!a\delta H\right) G_t
   \quad (t>1) ,
\label{Gtp1}
\end{equation}
where the initial evolution is set by
\begin{equation}
   G_1 =
   \left(1\!-\!\frac{aH_0}{2n}\right)^n
   U^\dagger_4
   \left(1\!-\!\frac{aH_0}{2n}\right)^n \, \delta_{\vec x,0} .
\label{G1}
\end{equation}
The kinetic energy operator $H_0$ is of $O(v^2)$, and
is given by
\begin{equation}
   H_0 = - { \Delta^{(2)} \over 2M_c^0 } ,
\label{H0}
\end{equation}
where $M_c^0$ is the bare quark mass and $\Delta^{(2)}$
is the lattice Laplacian.
Relativistic corrections are organized in powers of the heavy
quark velocity, with terms up to $O(v^6)$ considered here:
\begin{equation}
   \delta H = \delta H^{(4)} + \delta H^{(6)} .
\label{deltaH}
\end{equation}

Simulations with the complete set of $O(v^4)$ corrections 
were reported in Refs. \cite{Nups,Npsi}:
\begin{eqnarray}
   \delta H^{(4)} 
 & = &
   - c_1 { ( \Delta^{(2)} )^2 \over 8(M_c^0)^3 }
   + c_2 { ig \over 8 (M_c^0)^2 }
         ( \tilde {\bf \Delta} \cdot \tilde{\bf E}
         - \tilde {\bf E} \cdot \tilde {\bf \Delta} )
\nonumber \\
  & & 
   - c_3 { g \over 8( M_c^0 )^2 } 
         \mbox{{\boldmath$\sigma$}} \cdot 
         ( \tilde {\bf \Delta} \times \tilde {\bf E} 
         - \tilde {\bf E} \times \tilde {\bf \Delta} )
   - c_4 { g \over 2 M_c^0 } 
         \mbox{{\boldmath$\sigma$}} \cdot \tilde {\bf B}
\nonumber \\
  & &  
   + c_5 { a^2 \Delta^{(4)} \over 24 M_c^0 }
   - c_6 { a ( \Delta^{(2)} )^2 \over 16n (M_c^0)^2 } .
\label{H4}
\end{eqnarray}
The first two terms in $\delta H^{(4)}$ are spin-independent
relativistic corrections, and the last two terms
come from finite lattice spacing corrections to the lattice 
Laplacian and the lattice time derivative respectively.
The parameter $n$ is introduced to remove instabilities in the heavy 
quark propagator caused by the highest momentum modes of the 
theory \cite{N1992}.

While $\delta H^{(4)}$ yields spin-averaged spectra to 
next-to-leading order in the velocity expansion, it contains only 
leading order spin-dependent interactions. In potential model 
language the third term above ($c_3$) generates the 
spin-orbit and tensor potentials which drive the $P$-wave fine 
structure, while the fourth term ($c_4$) generates 
the color-magnetic $S$-wave hyperfine splitting.

Spin-dependent interactions at $O(v^6)$ were derived in 
Ref. \cite{N1992}:
\begin{eqnarray}
   \delta H^{(6)}
 & = &
   - c_7 { g \over 8 (M_c^0)^3 } 
         \left\{  \tilde \Delta^{(2)} , 
                  \mbox{{\boldmath$\sigma$}} \cdot \tilde {\bf B} 
         \right\} 
\nonumber \\
  & &
   - c_8 { 3g \over 64 (M_c^0)^4 } 
         \left\{  \tilde \Delta^{(2)} , 
                  \mbox{{\boldmath$\sigma$}} \cdot 
                ( \tilde {\bf \Delta} \times \tilde {\bf E} 
                - \tilde {\bf E} \times \tilde {\bf \Delta} )
         \right\} 
\nonumber \\
  & &
   - c_9 { i g^2 \over 8 (M_c^0)^3 } 
         \mbox{{\boldmath$\sigma$}} \cdot 
         \tilde {\bf E} \times \tilde {\bf E} .
\label{H6}
\end{eqnarray}
Notice the field strength bilinear ($c_9$), which is peculiar to 
the nonAbelian theory. There are additional $O(v^6)$ terms which 
contribute to spin-averaged spectra; these are not considered here.

The derivative operators and the fields are 
evaluated with their leading discretization errors removed, in order
to minimize the effects of lattice artifacts on the spin splittings.
This is indicated by the tilda superscripts
on these operators in Eqs. (\ref{H4}) and (\ref{H6}).
At leading order in $a$ the action of the symmetric lattice 
derivative $\Delta_i$ is defined by
\begin{equation}
   a \Delta_i G(x) \equiv 
   \case{1}{2} [ U_i(x) G(x + a\hat\imath) 
   - U^\dagger_i(x - a\hat\imath) G(x - a\hat\imath) ] .
\end{equation}
At tree level the $O(a^4)$-accurate derivative operator
$\tilde \Delta_i$ that is used in Eqs. (\ref{H4}) and (\ref{H6})
is given by \cite{N1992}
\begin{equation}
   \tilde \Delta_i = \Delta_i 
  - {a^2 \over 6} \Delta^{(+)}_i \Delta_i \Delta^{(-)}_i ,
\label{tDelta}
\end{equation}
where $\Delta^{(+)}_i$ and $\Delta^{(-)}_i$ are leading-order
forward and backward covariant finite differences:
\begin{eqnarray}
   a \Delta^{(+)}_i G(x) & \equiv & 
             U_i(x) G(x+a\hat\imath) - G(x) ,
\nonumber \\
   a \Delta^{(-)}_i G(x) & \equiv &     G(x) 
           - U^\dagger_i(x-a\hat\imath) G(x-a\hat\imath) .
\label{Deltapm}
\end{eqnarray}
At leading-order in $a$ the lattice Laplacian $\Delta^{(2)}$ 
is expressed in terms of covariant second order differences
\begin{equation}
    \Delta^{(2)} = \sum_i \Delta^{(2)}_i ,
\end{equation}
where
\begin{equation}
   a^2 \Delta^{(2)}_i G(x) = U_i(x) G(x+a\hat\imath)  - 2 G(x)
               + U^\dagger_i(x-a\hat\imath) G(x-a\hat\imath).
\end{equation}
The $O(a^4)$-accurate Laplacian $\tilde \Delta^{(2)}$ is
used in Eq. (\ref{H6}), and at tree-level is given by \cite{N1992}
\begin{equation}
   \tilde \Delta^{(2)} = \Delta^{(2)}
                       - {a^2 \over 12} \Delta^{(4)} ,
\end{equation}
where 
\begin{equation}
    \Delta^{(4)} = \sum_i \left( \Delta^{(2)}_i \right)^2 
\end{equation}
is a lattice representation of the continuum operator $\sum_i D_i^4$.
Note that discretization errors in $H_0$ have been removed in
this way by the addition of the $c_5$ term in Eq. (\ref{H4}).
What is new in the present work is the removal of discretization
errors in the leading order $P$-wave interaction ($c_3$),
and the Darwin term ($c_2$), as well as the addition of the 
$O(v^6)$ spin terms in Eq. (\ref{H6}).

To complete the removal of discretization errors from 
the spin-dependent interactions
$O(a^4)$-accurate chromo-electric and -magnetic fields 
have been used in these simulations. In Refs. \cite{Nups,Npsi}
the leading order field strength $F_{\mu\nu}$ was evaluated
using the standard traceless clover operator: 
\begin{equation}
   F_{\mu\nu}(x) = {1 \over 2i} 
        \left( \Omega_{\mu\nu}(x) - \Omega_{\mu\nu}^\dagger(x) \right)
   - {1 \over 3} \mbox{Im} \left( \mbox{Tr} \, \Omega_{\mu\nu}(x) \right) ,
\end{equation}
where $\Omega_{\mu\nu}$ is an average over the 4 counterclockwise
plaquettes in the $(\mu,\nu)$ plane containing the site $x$
\begin{equation}
   \Omega_{\mu\nu} = - {1 \over 4}
   \sum_{\alpha=\pm\mu} \sum_{\beta=\pm\nu} 
   U_\alpha(x) U_\beta(x+\hat\alpha) 
   U_{-\alpha}(x+\hat\alpha+\hat\beta)
            U_{-\beta}(x+\hat\beta) .
\end{equation}
The $O(a^4)$-accurate field strength $\tilde F_{\mu\nu}$
used here is computed following the analysis of Ref. \cite{N1992}:
\begin{eqnarray}
   \tilde F_{\mu\nu}(x) & = & {5 \over 3} F_{\mu\nu}(x)
   - {1 \over 6} \Bigl[ 
     U_\mu(x) F_{\mu\nu}(x+\hat\mu) U_\mu^\dagger(x)
\nonumber \\
   & & + U_\mu^\dagger(x-\hat\mu) F_{\mu\nu}(x-\hat\mu) 
                                  U_\mu(x-\hat\mu)
       - (\mu \leftrightarrow \nu) \Bigr] .
\end{eqnarray}
The ``improved'' chromo-electric and -magnetic fields are 
defined in terms of the field strength,
$\tilde E_i = \tilde F_{4i}$ and 
$\tilde B_i = \frac12 \epsilon_{ijk} \tilde F_{jk}$.

It should be noted that some of the discretization errors 
that are removed by using $\tilde\Delta$, $\tilde\Delta^{(2)}$,
and $\tilde F_{\mu\nu}$ in Eqs. (\ref{H4}) and (\ref{H6}), in place 
of their leading order counterparts, may in fact be comparable 
to or smaller than higher order relativistic corrections 
that are not included here. This includes for example the
use of $\tilde\Delta^{(2)}$ and $\tilde F_{\mu\nu}$
in the $O(v^6)$ terms in Eq. (\ref{H6}), and the use 
of the improved operators in the Darwin term ($c_2$). 
On the other hand, the use of improved operators in the 
leading spin-dependent interactions ($c_3$ and $c_4$) 
corrects for errors of $O(a^2 M_c v^4)$ in the spin splittings which, 
for the range of lattice spacings studied here, may be 
comparable to the $O(M_c v^6)$ contributions from $\delta H^{(6)}$.

At tree-level all of the coefficients $c_i$ in Eqs. (\ref{H4}) 
and (\ref{H6}) are one. However, very large radiative corrections 
in the lattice theory can arise from tadpoles that are induced by
the nonlinear connection between the link variables $U_\mu$
and the continuum gauge fields. Most of the effects of tadpoles
can be removed by a mean-field renormalization of the 
link \cite{LepMac}:
\begin{equation}
   U_\mu(x) \rightarrow {U_\mu(x) \over u_0} .
\label{u0}
\end{equation}
The links are rescaled in the simulation before they
are input to the quark propagator subroutine, to be sure
that Eq. (\ref{u0}) is correctly implemented in all
terms in the heavy quark action.

In most previous work the fourth root of the average 
plaquette has been used to set the value of $u_0$:
\begin{equation}
   u_{0,P} \equiv
       \left\langle \case13 \mbox{ReTr} \, U_{\mbox{pl}} 
       \right\rangle^{1/4} .
\label{uplaq}
\end{equation}
Simulations were done here with this renormalization prescription.
In addition, simulations were also done using
the mean link in Landau gauge to set $u_0$, as 
recently suggested by Lepage \cite{LepLandau}
\begin{equation}
   u_{0,L} \equiv
       \left\langle \case13 \mbox{ReTr} \, U_\mu \right\rangle, 
       \quad \partial_\mu A_\mu = 0 ,
\label{ulandau}
\end{equation}
where a standard lattice implementation of the continuum Landau
gauge fixing is used \cite{FFTLandau} (it was found
that the removal of leading discretization errors in the lattice
version of $\partial_\mu A_\mu = 0$ results in a negligible change
to the value of $u_{0,L}$).

Finally, the gauge-field configurations were generated using an
$O(a^4)$-accurate tadpole-improved action \cite{LepCoarse}
\begin{equation}
    S[U] = \beta \sum_{\mbox{pl}} \case13
           \mbox{ReTr} \left(1 - U_{\mbox{pl}}\right)
         - {\beta \over 20 u_0^2} \sum_{\mbox{rt}} \case13
           \mbox{ReTr} \left(1 - U_{\mbox{rt}}\right) ,
\label{Sglue}
\end{equation}
where the sums are over all oriented $1\times1$ plaquettes
and $1\times2$ rectangles.

\section{Meson propagators}
In order to increase the overlap of the meson propagators with
the ground states of interest here, a gauge-covariant
smearing procedure has been used \cite{LepSmear}.
A meson creation operator is constructed from quark 
and antiquark creation operators $\psi^\dagger$ and 
$\chi^\dagger$ \cite{LepThac,Nups,Npsi}:
\begin{equation}
   \sum_{\vec x} \psi^\dagger(\vec x) \Gamma(\vec x) 
                 \chi^\dagger(\vec x) ,
\end{equation}
with
\begin{equation}
   \Gamma(\vec x) \equiv \Omega(\vec x) \gamma(\vec x) ,
\end{equation}
where $\Omega(\vec x)$ is a $2 \times 2$ matrix in spin-space,
with derivative operator entries, which gives the quantum
numbers of the state of interest. $\gamma(\vec x)$ is a
gauge-covariant local smearing operator, which is taken
to have the simple form \cite{LepSmear}
\begin{equation}
   \gamma(\vec x) = \left( 1 + \epsilon \Delta^{(2)}(\vec x)
                   \right)^{n_s} 
\label{gamma}
\end{equation}
(an invariant under the lattice cubic group).
The weight $\epsilon$ and the number of smearing iterations $n_s$
are adjusted to optimize the overlap with the ground state.

The meson correlation function $G_{\rm meson}$ at zero
momentum is then given by
\begin{equation}
   G_{\rm meson}(\vec p = 0, t) = 
   \sum_{\vec y} \mbox{Tr} \left[ 
   G_t^\dagger(\vec y - \vec x) \Gamma_{(sk)}^\dagger(\vec y)
   G_t(\vec y - \vec x) \Gamma_{(sc)}(\vec x) \right] ,
\end{equation}
where different smearing parameters may be used at the source and
sink, and where a single spatial origin $\vec x$ for the meson
propagator was generally used.
Finite momentum propagators for the ${}^1S_0$ were analyzed
using a local $\delta$-function source and sink:
\begin{equation}
   G_{\rm meson}(\vec p,t) = 
   \sum_{\vec y} \mbox{Tr} \left[
   G_t^\dagger(\vec y - \vec x) G_t(\vec y - \vec x) \right] 
   e^{-i \vec p \cdot (\vec y - \vec x)} .
\end{equation}

Correlation functions were computed for the 
${}^1S_0$ ($\Omega = I$), 
${}^3S_1$ ($\Omega = \sigma_i$) and 
${}^1P_1$ ($\Omega = \Delta_i$) mesons. 
The three triplet $P$-wave correlators 
(${}^3P_0$, ${}^3P_1$, ${}^3P_2$) were also analyzed; the
relevant operators $\Omega$ for these states are tabulated
in Ref. \cite{Nups}. Only selected meson polarizations were used.
Propagators were generated for all (equal) quark-antiquark 
colors but, in order to save computer time, 
the initial quark and antiquark spins were set to 1. 
The ${}^3S_1 z$ and ${}^1S_0$ states were thus obtained from 
a single propagator, since the $1\times 1$ component of the 
spin matrix $\Omega$ is the same for both states \cite{Nups}.
All three polarizations of the ${}^1P_1$ were generated, but 
only one from each of the ${}^3P_1$ and ${}^3P_2$.

\section{Results}
Three lattices were generated using the mean link in Landau
gauge to set the tadpole factor ($u_{0,L}$)
and four lattices with comparable spacings
were generated using the average plaquette tadpole ($u_{0,P}$).
The parameters of these seven lattices are given in Tables 
\ref{betaL} and \ref{betaP}. 
In order to distinguish between
the two sets of simulations, $\beta_L$ is used to denote
the lattice coupling when Landau-gauge tadpole renormalization is used,
and $\beta_P$ when the average plaquette is used.

A standard Cabbibo-Marinari
pseudo-heat bath was used to generated the gauge field
configurations. Integrated autocorrelation 
times $\tau_{\rm int}$ were checked for all correlation functions,
and were found to be remarkably short; 10 updates between
measurements yields $\tau_{\rm int} \alt 0.5$ on the three lattices
with $a \lesssim .2$~fm, and on the coarser lattices
5 updates was found to be sufficient.

Smeared-smeared correlators were used for the $P$-waves,
while local sources and smeared sinks were used for the $S$-waves.
Ten smearing iterations [$n_s=10$ in Eq. (\ref{gamma})] 
were used for the three lattices with $a \lesssim .2$~fm,
5 iterations for the lattices with $a$ near .28~fm, 
and 2 iterations for the lattices with $a \approx .39$~fm. 
A smearing weight $\epsilon = 1/12$ was used in all cases.

The lattice spacings are determined from the spin-averaged
$1P - 1S$ mass difference, following Refs. \cite{Nups,Npsi}.
This mass difference is known to be independent of the 
quark mass in the charm to bottom region.
For this purpose the singlet ${}^1P_1$ and the 
spin-averaged ${}^3S_1$, ${}^1S_0$ masses were used.
The simulation results for the splitting were fixed to
the experimental value for charmonium of 458~MeV. 

After the lattice spacing was extracted, the kinetic
mass $M_{\rm kin}$ (in physical units) of the ${}^1S_0$ state was 
determined by fitting the energy $E_{\bf P}$ of the 
boosted state to the form
\begin{equation}
   E_{\bf P} - E_0 = { {\bf P}^2 \over 2 M_{\rm kin} } .
\label{Ep}
\end{equation}
Fits were made to the state with momentum components $(1,0,0)$
in units of $2\pi/(Na)$; in some cases simultaneous fits including
states with momentum components $(1,1,0)$ and $(1,1,1)$
were also done, with little change to the fit values of
$M_{\rm kin}$. A dispersion relation including
relativistic corrections \cite{Nups,Npsi} was also tried,
and the resulting changes to the fit values of $M_{\rm kin}$ were
within a few percent, as expected on these lattices.

The correct values of the bare charm mass $M_c^0$ were 
determined by tuning so that $M_{\rm kin}$ agrees with the 
experimental value of the mass
of the $\eta_c$ (2.98~GeV). The bare masses are listed in 
Tables \ref{betaL} and \ref{betaP}, and all yield 
$M_{\rm kin} = 2.9(1)$~GeV.
For bare masses $a M_c^0 < 1.2$ a stability parameter
$n = 4$ was used in the quark propagators, Eqs. (\ref{Gtp1}) and 
(\ref{G1}); for $1.2 < a M_c^0 < 1.5$ $n=3$ was used, and
for the larger bare masses $n = 2$ was used.

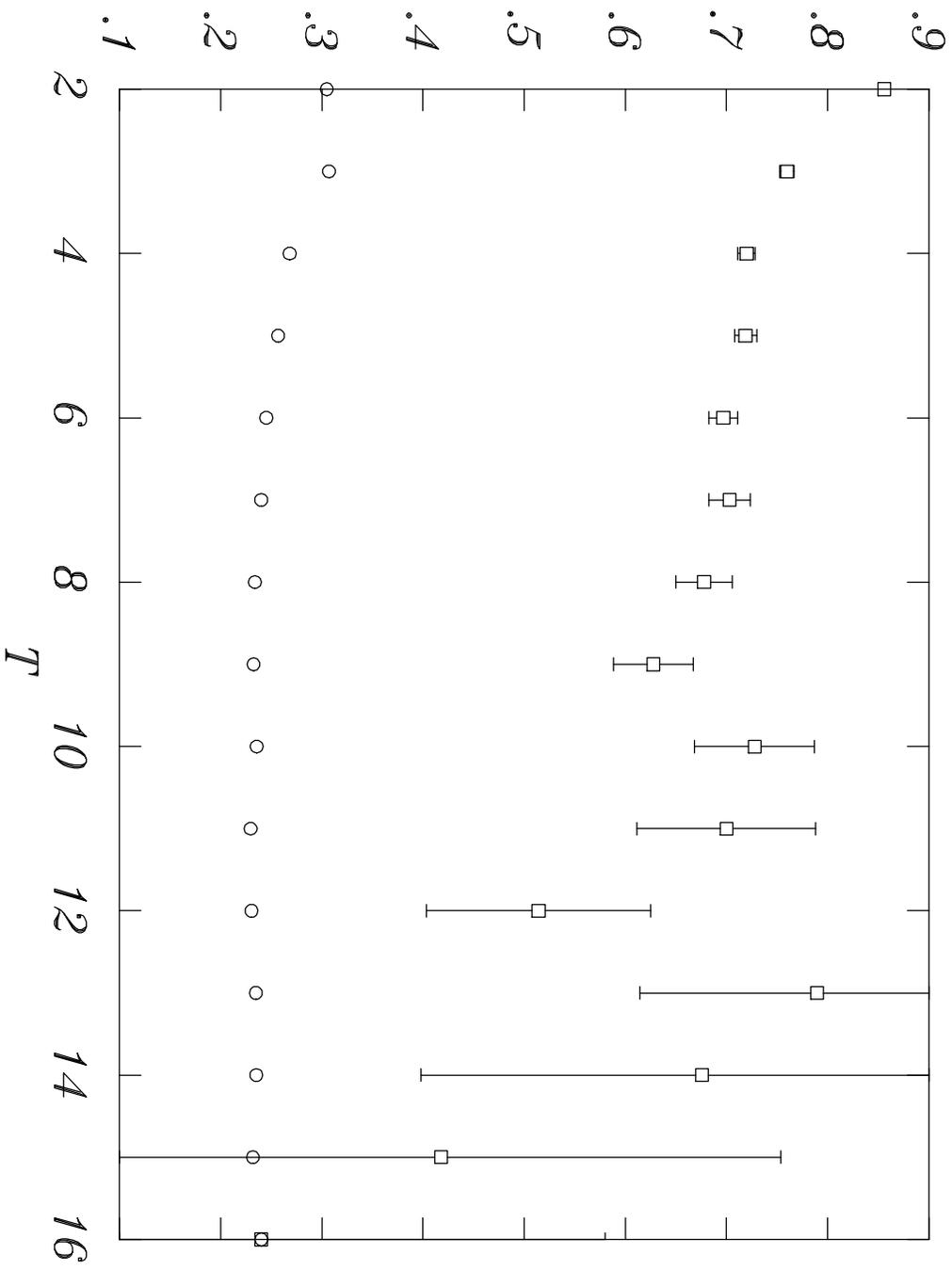
\begin{figure}
\caption{Effective mass plot for $\beta_L=7.4$ ($a=.18$~fm):
${}^1P_1$ state ($\Box$) and ${}^1S_0$ state ($\circ$).}
\label{FL74}
\end{figure}

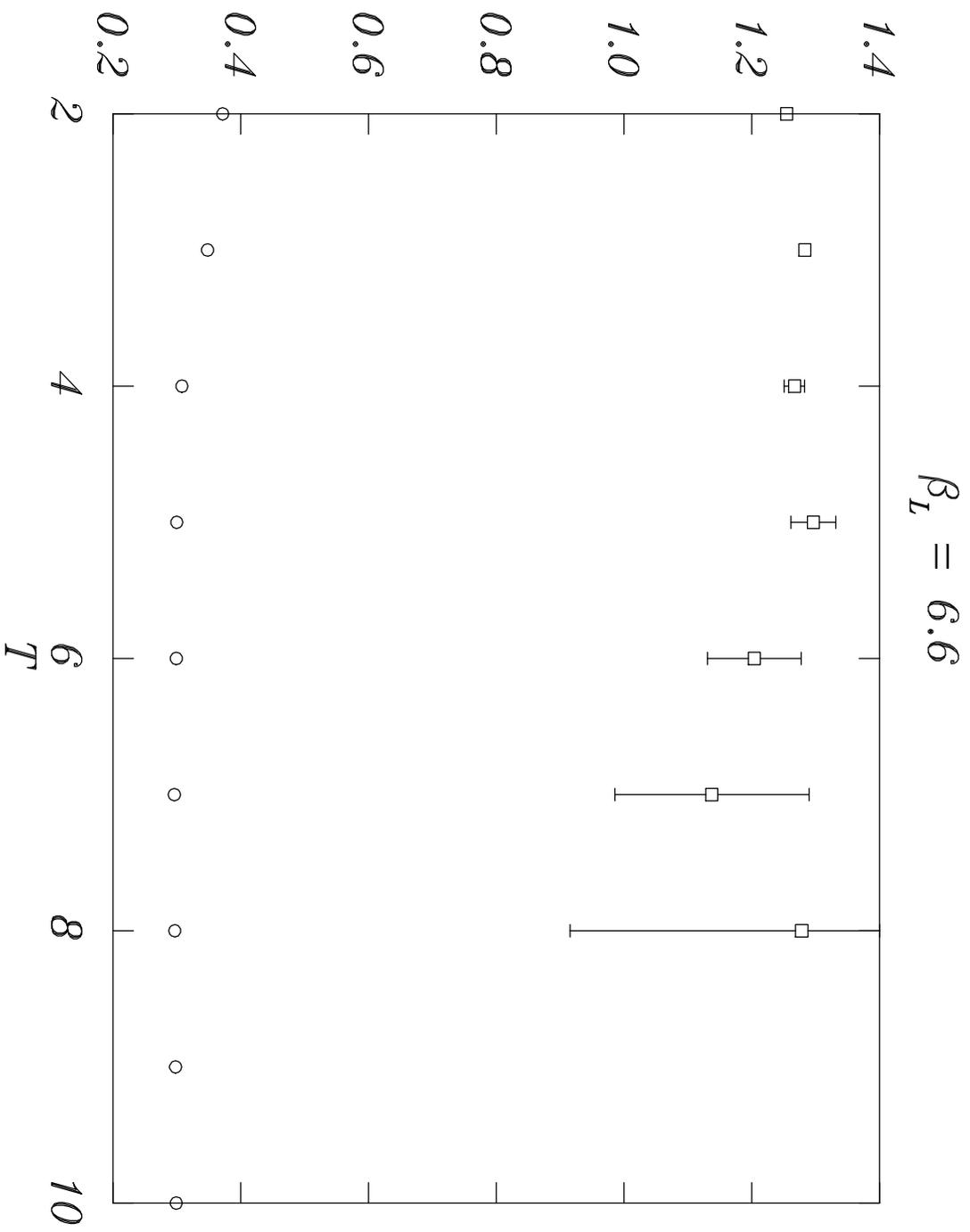
\begin{figure}
\caption{Effective mass plot for $\beta_L=6.6$ ($a=.39$~fm):
${}^1P_1$ state ($\Box$) and ${}^1S_0$ state ($\circ$).}
\label{FL66}
\end{figure}

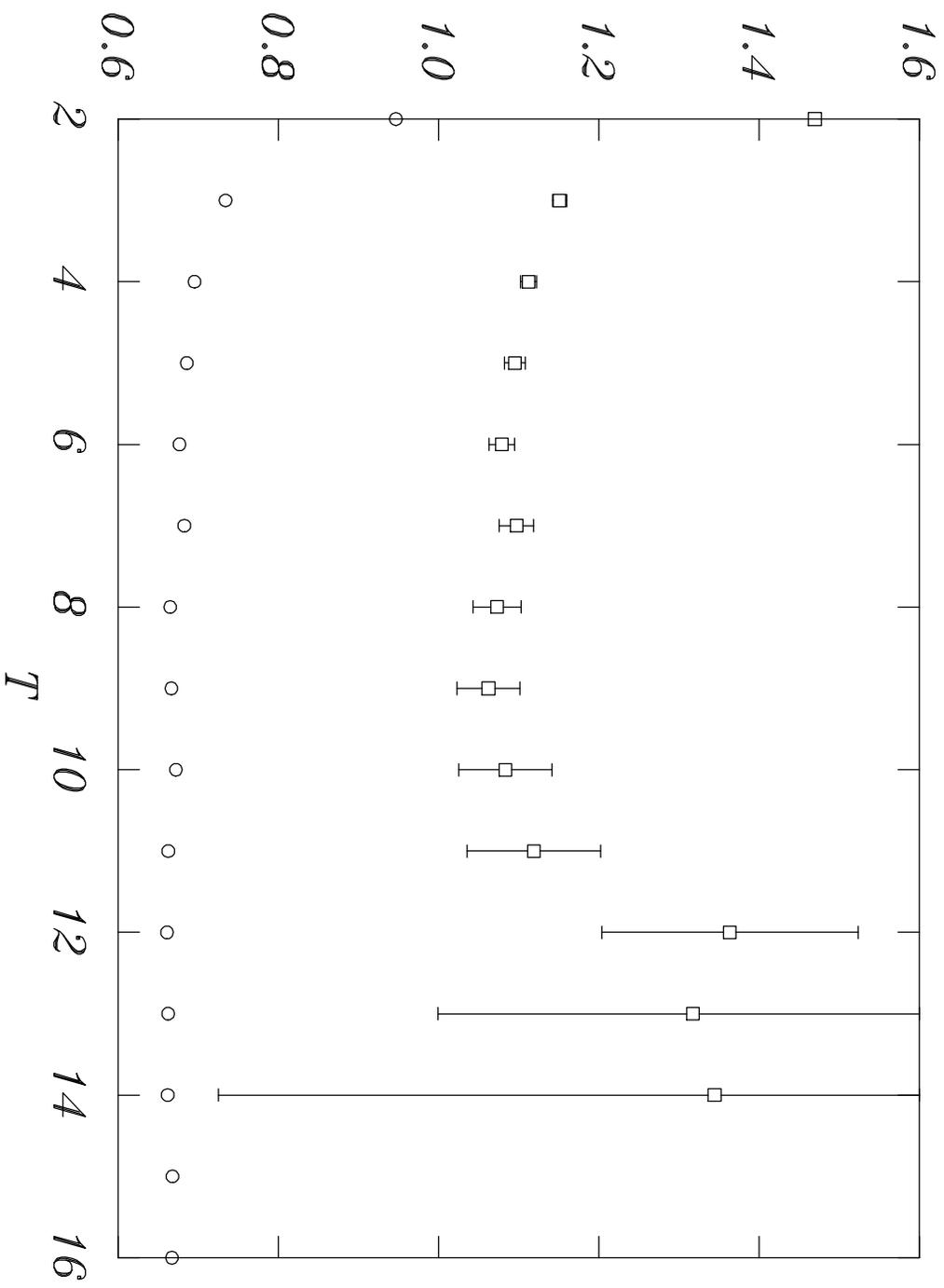
\begin{figure}
\caption{Effective mass plot for $\beta_P=7.2$ ($a=.17$~fm):
${}^1P_1$ state ($\Box$) and ${}^1S_0$ state ($\circ$).}
\label{FP72}
\end{figure}

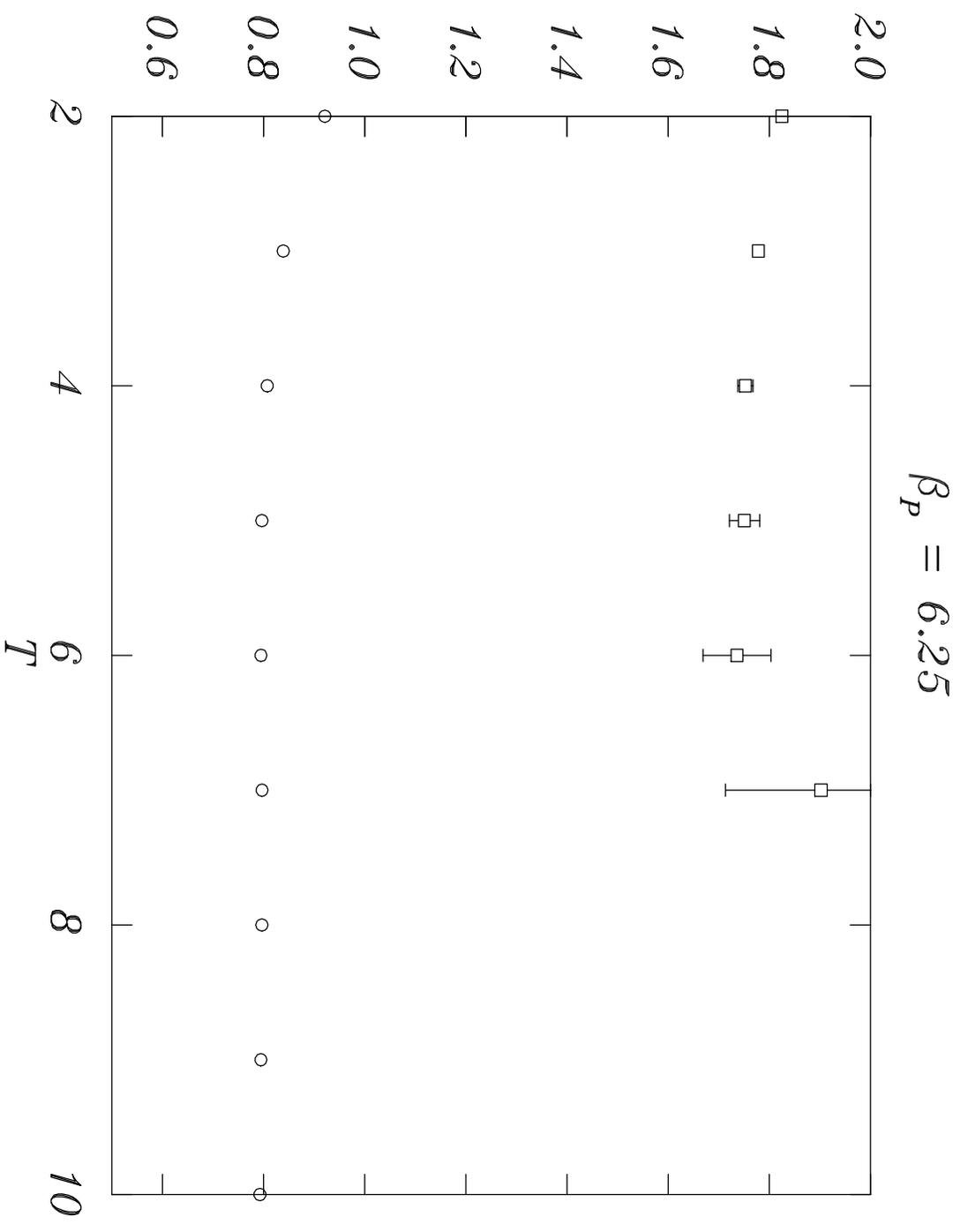
\begin{figure}
\caption{Effective mass plot for $\beta_P=6.25$ ($a=.39$~fm):
${}^1P_1$ state ($\Box$) and ${}^1S_0$ state ($\circ$).}
\label{FP625}
\end{figure}

Effective mass plots $m_{\rm eff}(T) = -\log(G(T)/G(T-1))$ for 
several lattices are shown in Figures \ref{FL74}--\ref{FP625},
using jackknife errors.
Single exponential fits to the correlation functions are
used to get the best estimates of the masses
of the individual states. The fitting 
procedure included the full covariance matrix for the data,
using the svd algorithm \cite{Richard}. The correlation
functions for states of a given partial wave are highly
correlated; following Refs. \cite{Nups,Npsi} a spin
splitting $\delta E$ was obtained from a correlated
fit of the form
\begin{eqnarray}
   G_{\rm meson,A}(t) & = & c_A e^{-E_A t} ,
\nonumber \\
   G_{\rm meson,B}(t) & = & c_B e^{-(E_A + \delta E) t} .
\label{GdeltaE}
\end{eqnarray}

Detailed fit results for several lattices are reported in 
Tables \ref{TL74}--\ref{TP625}. The statistical errors were 
estimated using bootstrap ensembles of 1000 samples.
Final estimates of the dimensionless energies
are obtained from these fits by finding two or three successive
$t_{\rm min}/t_{\rm max}$ intervals for which the fit
results overlap within statistical errors;
acceptable $Q$ values were obtained in all cases at
these $t_{\rm min}/t_{\rm max}$ values.
Estimates of the systematic errors in the final fit
results are taken from the largest statistical errors
in the overlapping intervals. 

The final fit results are 
shown in Tables~\ref{LFinal} and \ref{PFinal},
where the resulting lattice spacings and hyperfine splittings
in physical units are also given. The dominant error
in the splitting comes from the systematic error in the determination
of the bare quark mass. The error in the mass comes in part from
the uncertainty in $a^{-1}$, which has been included in
the error estimates for the splittings in physical units; 
however, there is a further systematic error 
of order 10\% in the quark mass determination, coming from 
higher order (spin-independent) relativistic corrections \cite{Npsi}.

\section{Discussion}
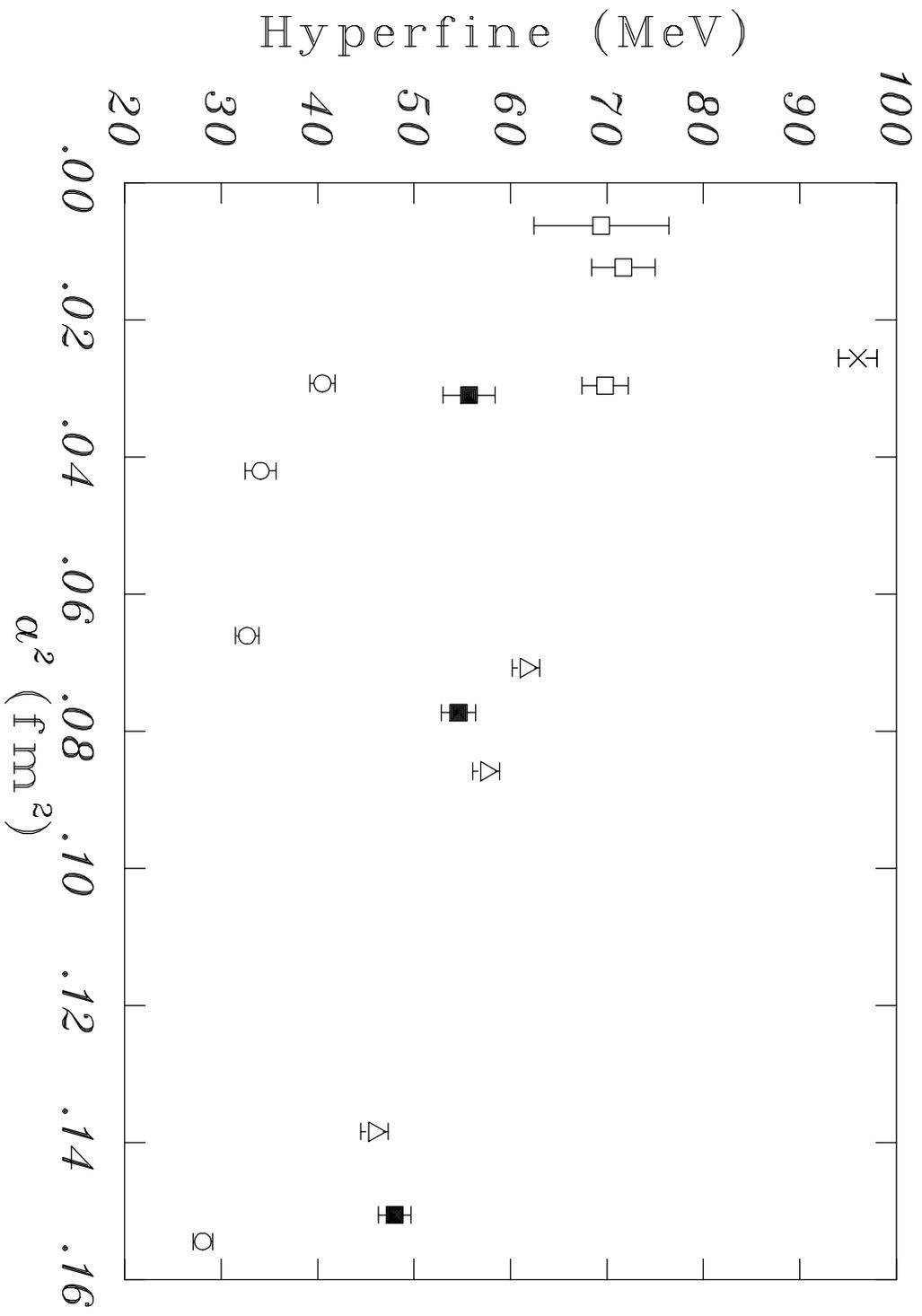
\begin{figure}
\caption{Hyperfine splittings versus lattice spacing squared.
The next-to-leading order NRQCD results with Landau gauge
tadpoles (\protect\rule{2mm}{2mm}) and with average plaquette
tadpoles ($\circ$) are shown, as well as the leading-order
result ($\times$) from Ref. \protect\cite{Npsi}. Also shown are
results obtained with the relativistic Fermilab action
($\Box$) \protect\cite{FNAL}, and coarse lattice results
from the relativistic tadpole-improved D234 action
($\triangle$) \protect\cite{D234}.
The experimental value is $(118 \pm 2)$~MeV.}
\label{FigHyper}
\end{figure}

The hyperfine splittings are plotted as a function
of lattice spacing in Fig. \ref{FigHyper}, where the 
results from the relativistic Fermilab action \cite{FNAL} 
and the leading order NRQCD calculation \cite{Npsi} are included.
Some coarse lattice results from the tadpole-improved 
relativistic D234 \cite{D234} action are also shown.

The next-to-leading order corrections result
in a very large reduction in the hyperfine splitting, down
by about 60\% from the leading order result on a lattice
of comparable spacing, when the same plaquette tadpole
renormalization scheme ($u_{0,P}$) is used in both cases.

The hyperfine splitting shows very little $a$ dependence when 
the Landau gauge tadpole scheme $u_{0,L}$ is used.
The results with $u_{0,P}$ on the other hand have large 
discretization errors, which prevents a reliable 
extrapolation to zero lattice spacing in this case; however,
the results are not inconsistent with an extrapolation to the
same splitting as is obtained with Landau gauge $u_{0,L}$.

From these results the hyperfine splitting at next-to-leading
order in the velocity expansion, and at zero lattice spacing,
can be estimated at roughly $(55\pm5)$~MeV. This can be
compared with the Fermilab action result of approximately 
$(70\pm3)$~MeV \cite{FNAL}. While the two calculations have
different systematic errors, this suggests that further 
relativistic corrections, beyond the next-to-leading-order 
considered here, are again very large. [The experimental value 
is $(118 \pm 2)$~MeV, which indicates that there are significant 
effects due to quenching \cite{FNAL,Npsi}.]

These considerations are supported by results obtained here for
the triplet $P$-wave spectra (${}^3P_0$, ${}^3P_1$, ${}^3P_2$).
The next-to-leading order fine structure splittings are
apparently very small, and much better statistics
are required for an accurate determination.
Simulations with $u_{0,L}$ give a 
${}^3P_2 - {}^3P_0$ splitting of about $(30\pm15)$~MeV,
down by about 60\% from the leading-order result
of $(110 \pm 10)$~MeV reported in Ref. \cite{Npsi}
(the experimental value is $(141 \pm 10)$~MeV).
The next-to-leading order fine structure splittings
with $u_{0,P}$ are even harder to measure,
but the results suggest that the splittings
may actually be in the wrong order with that tadpole scheme
(as least in the range of lattice spacings analyzed here).

These results indicate that the NRQCD velocity expansion for
charmonium may be poorly convergent, with the first three terms in 
the expansion for the hyperfine splitting apparently 
oscillating in sign. However there are other sources of 
systematic error in the NRQCD action which must also be
considered. Radiative corrections to the 
operator coefficients $c_3$ and $c_4$ in Eq. (\ref{H4})
are of particular importance. 

It is worthwhile to assess the relative importance
of the various next-to-leading order corrections that have
been considered here. Within a given tadpole renormalization 
scheme the most important correction for the charmonium system
comes from the relativistic spin-dependent interactions
$\delta H^{(6)}$ [Eq. (\ref{H6})], and these drive
the large reduction in the splittings. There is some 
indication that the field strength bilinear (term $c_9$)
plays a relatively small role in these effects.
The use of $O(a^4)$-accurate clover fields increases the
spin splittings, a correction amounting to about 20\% of the 
$O(a^2)$-accurate splittings on the coarsest lattices considered here 
(this correction falls below about 10\% at the smallest spacings).

The effect of a change in the tadpole renormalization scheme is
very significant. The spin splittings vary as $1/u_0^4$
due to the renormalization of the clover field interactions (except
for the field strength bilinear, which varies
as $1/u_0^8$). This renormalization causes most of the change
due to tadpole scheme, as can be seen from the values of
$u_{0,L}$ and $u_{0,P}$ in Tables \ref{betaL} and
\ref{betaP} (additional changes in the splittings are presumably 
caused by the renormalization of the gluonic action).
For example, at $a \approx .18$~fm the ratio $(u_{0,P} / u_{0,L})^4$
is about 1.2, and at $a \approx .39$~fm the ratio is about 1.5.

Relativistic corrections are expected to be much smaller for the 
Upsilon system. Preliminary results from a next-to-leading order 
calculation at $\beta_P = 7.2$ give an $\Upsilon - \eta_b$ hyperfine 
splitting of $(22.4 \pm 1.3)$~MeV, using a bare mass $a M_b^0=3.15$.
This is within errors of the leading-order result reported in 
Ref. \cite{Nups57} on a lattice with comparable spacing 
(using $u_{0,P}$), which is consistent with the velocity expansion,
given the fact that $v_\Upsilon^2 \approx 0.1$.
The small size of the net correction, relative to the charmonium
system, also appears to be driven by a near cancellation of the
$O(a^2 M_Q v^4)$ discretization corrections (which tend to raise 
the splittings) and the $O(M_Q v^6)$ relativistic corrections
(which tend to lower them), which are much closer in magnitude
in the Upsilon system. This was demontrated by doing a calculation
with a Wilson gauge field action at $\beta=5.7$, without removing 
discretization errors in the clover field, but including the 
$O(M_Q v^6)$ relativistic interactions; this results in a reduction
of the Upsilon hyperfine splitting by about 15\% compared to the 
leading order calculation \cite{Nups57} at the same $\beta$.

It is also interesting to note that the lattice spacing as
determined from the $1P - 1S$ splitting is different for
Upsilon and charmonium. At $\beta_P=7.2$ preliminary results
give $a_\Upsilon = 0.146(9)$~fm, compared to
$a_\psi = 0.171(4)$~fm. This is comparable to the difference 
between the two determinations of the spacing in Ref. \cite{Nups57}.

\section{Summary and Outlook}
It has been shown that spin-dependent interactions 
at next-to-leading order in the NRQCD velocity expansion 
yield very large corrections to the charmonium spin splittings,
down by about 60\% from their values at leading order
(when the same tadpole-improvement scheme is used on
lattices with comparable spacings). There are indications that 
further relativistic corrections for charmonium are also very
large. The corrections to the hyperfine splitting in the Upsilon 
system are small. More work needs to be done in order to assess the 
validity of the NRQCD effective action in simulations of charmonium,
including better measurements of the triplet $P$-wave spectra.
Estimates of the radiative corrections to the operator
coefficients in the heavy quark action are also needed. More
complete calculations of the Upsilon splittings would also
provide useful information. 
The results obtained here provide further support for 
the use of Landau gauge tadpole renormalization.

\acknowledgments

I am indebted to G. P. Lepage and R. M. Woloshyn
for many helpful discussions and suggestions.
I also thank C. Davies, J. Shigemitsu 
and J. Sloan for useful conversations.
This work was supported in part by the 
Natural Sciences and Engineering Research Council of Canada.

\begin{table}
\begin{center}
\begin{tabular}{ccccccc}
  $\beta_L$   
& $\langle \case13 \mbox{ReTr} \, U_\mu \rangle$
& $\langle \case13 \mbox{ReTr} \, U_{\mbox{pl}} \rangle^{1/4}$
& $a$ (fm)    & $a M_c^0$   & Volume  & $N_{\rm meas}$ \\
\tableline
7.4 &  .829   &  .875   & .18  & 1.18  & $10^3 \times 16$ & \ 752 \\
7.0 &  .780   &  .850   & .28  & 1.90  & $6^3 \times 10$  & 1740  \\
6.6 &  .743   &  .825   & .39  & 2.65  & $6^3 \times 10$  & 2092  \\
\end{tabular}
\end{center}
\caption{Simulation parameters using the Landau gauge mean
link to determine the tadpole renormalization, 
$u_{0,L} = \langle \case13 \mbox{ReTr} \, U_\mu \rangle$
(second column). $N_{\rm meas}$ is the number of
configurations used for propagator measurements.}
\label{betaL}
\end{table}

\begin{table}
\begin{center}
\begin{tabular}{ccccccc}
  $\beta_P$   
& $\langle \case13 \mbox{ReTr} \, U_\mu \rangle$
& $\langle \case13 \mbox{ReTr} \, U_{\mbox{pl}} \rangle^{1/4}$
& $a$ (fm)    & $a M_c^0$  & Volume   & $N_{\rm meas}$ \\
\hline
7.2    &  .834   &  .874   & .17   & 0.81  & $10^3 \times 16$ & \ 474 \\
7.0    &  .810   &  .865   & .21   & 1.10  & $8^3 \times 10$  & \ 923 \\
6.8    &  .786   &  .854   & .26   & 1.43  & $6^3 \times 10$   & 1815 \\
6.25   &  .738   &  .821   & .39   & 2.30  & $6^3 \times 10$   & 2841 \\
\end{tabular}
\end{center}
\caption{Simulation parameters using the average plaquette
to determine the tadpole renormalization, 
$u_{0,P}=\langle \case13 \mbox{ReTr} \, U_{\mbox{pl}} \rangle^{1/4}$
(third column).}
\label{betaP}
\end{table}

\begin{table}
\begin{center}
\begin{tabular}{ccccc}
$t_{\rm min}/t_{\rm max}$  & ${}^1P_1$     & ${}^3S_1$    & ${}^1S_0$ 
& ${}^3S_1 - {}^1S_0$        \\
\hline
2/16  & 0.735(5)\ \   & 0.298(2)  & 0.242(1)  & 0.0537(6)\ \  \\
3/16  & 0.715(6)\ \   & 0.293(2)  & 0.239(1)  & 0.0533(5)\ \  \\
4/16  & 0.709(9)\ \   & 0.289(2)  & 0.237(1)  & 0.0518(6)\ \  \\
5/16  & 0.694(12)     & 0.286(2)  & 0.235(1)  & 0.0506(6)\ \  \\
6/16  & 0.688(16)     & 0.285(2)  & 0.235(1)  & 0.0497(7)\ \  \\
7/16  & 0.672(23)     & 0.284(2)  & 0.234(1)  & 0.0500(8)\ \  \\
8/16  & 0.661(30)     & 0.284(2)  & 0.234(1)  & 0.0498(9)\ \  \\
9/16  & 0.700(51)     & 0.284(2)  & 0.234(1)  & 0.0494(10)    \\
\end{tabular}
\end{center}
\caption{Examples of fits to Landau-gauge tadpole simulation 
at $\beta_L=7.4$ ($a = .18$~fm).
Single exponential fits were used for each individual state, 
and a correlated $\delta E$ fit for the ${}^3S_1$ and ${}^1S_0$ 
was used to get the hyperfine splitting.}
\label{TL74}
\end{table}

\begin{table}
\begin{center}
\begin{tabular}{ccccc}
$t_{\rm min}/t_{\rm max}$  & ${}^1P_1$     & ${}^3S_1$    & ${}^1S_0$ 
& ${}^3S_1 - {}^1S_0$        \\
\hline
2/10  &  1.280(8)\ \    &  0.404(1)  &  0.304(1)  &  0.0980(6)\ \   \\
3/10  &  1.268(15)      &  0.395(1)  &  0.299(1)  &  0.0969(6)\ \   \\
4/10  &  1.279(35)      &  0.393(1)  &  0.298(1)  &  0.0950(7)\ \   \\
5/10  &  1.198(71)      &  0.392(2)  &  0.298(1)  &  0.0945(7)\ \   \\ 
6/10  &  1.18(14)\ \ \  &  0.392(2)  &  0.298(1)  &  0.0946(10)     \\
7/10  &  1.44(43)\ \ \  &  0.392(2)  &  0.298(1)  &  0.0944(12)     \\
\end{tabular}
\end{center}
\caption{Examples of fits to Landau-gauge tadpole simulation 
at $\beta_L=6.6$ ($a = .39$~fm).
Fits were done as in Table~\protect\ref{TL74}.}
\label{TL66}
\end{table}

\begin{table}
\begin{center}
\begin{tabular}{ccccc}
$t_{\rm min}/t_{\rm max}$   & ${}^1P_1$     & ${}^3S_1$    & ${}^1S_0$ 
& ${}^3S_1 - {}^1S_0$        \\
\hline
2/16  &  1.120(6)\ \  &  0.715(2)  &  0.675(2)  &  0.0399(5) \\
3/16  &  1.101(7)\ \  &  0.708(2)  &  0.670(2)  &  0.0395(5) \\
4/16  &  1.089(9)\ \  &  0.706(2)  &  0.669(2)  &  0.0369(5) \\
5/16  &  1.083(12)    &  0.704(2)  &  0.667(2)  &  0.0363(6) \\
6/16  &  1.087(15)    &  0.702(2)  &  0.667(2)  &  0.0357(7) \\
7/16  &  1.083(21)    &  0.700(2)  &  0.666(2)  &  0.0352(7) \\
8/16  &  1.093(29)    &  0.701(2)  &  0.666(2)  &  0.0352(8) \\
\end{tabular}
\end{center}
\caption{Examples of fits to average plaquette tadpole simulation 
at $\beta_P=7.2$ ($a = .17$~fm).
Fits were done as in Table~\protect\ref{TL74}.}
\label{TP72}
\end{table}

\begin{table}
\begin{center}
\begin{tabular}{ccccc}
$t_{\rm min}/t_{\rm max}$   & ${}^1P_1$     & ${}^3S_1$    & ${}^1S_0$ 
& ${}^3S_1 - {}^1S_0$        \\
\hline
2/10 &  1.773(7)\ \  &  0.858(1)  &  0.801(1)  &  0.0575(3)  \\
3/10 &  1.754(14)    &  0.853(1)  &  0.797(1)  &  0.0570(3)  \\ 
4/10 &  1.754(26)    &  0.852(1)  &  0.795(1)  &  0.0565(3)  \\
5/10 &  1.771(60)    &  0.851(1)  &  0.795(1)  &  0.0561(4)  \\
6/10 &               &  0.851(1)  &  0.795(1)  &  0.0560(4)  \\
7/10 &               &  0.850(1)  &  0.795(1)  &  0.0556(6)  \\
\end{tabular}
\end{center}
\caption{Examples of fits to average plaquette tadpole simulation 
at $\beta_P=6.25$ ($a = .39$~fm).
Fits were done as in Table~\protect\ref{TL74}.}
\label{TP625}
\end{table}

\begin{table}
\begin{center}
\begin{tabular}{cccccc}
$\beta_L$     & ${}^1P_1$   & ${}^1S_0$  & ${}^3S_1 - {}^1S_0$   
       &  $a$   & Hyperfine  \\
& & &  & (fm)   &   (MeV)  \\
\hline
7.4\ \ & 0.68(2) & 0.234(1) & 0.0497(7) & 0.176(9)\ \ & 55.7(27)  \\ 
7.0\ \ & 1.01(2) & 0.305(1) & 0.0772(8) & 0.278(9)\ \ & 54.6(18)  \\
6.6\ \ & 1.27(3) & 0.298(1) & 0.0945(7) & 0.388(13)   & 48.0(17)  \\
\end{tabular}
\end{center}
\caption{Final fit results for the dimensionless energies from
Landau-gauge tadpole simulations; the resulting lattice spacings
and hyperfine splittings in physical units are also shown.
The quoted errors in the hyperfine splittings in physical
units include the systematic errors in $a$.}
\label{LFinal}
\end{table}

\begin{table}
\begin{center}
\begin{tabular}{cccccc}
$\beta_P$     & ${}^1P_1$   & ${}^1S_0$  & ${}^3S_1 - {}^1S_0$   
       &  $a$   & Hyperfine  \\
& & &  & (fm)   &   (MeV)  \\
\hline
7.2\ \ & 1.09(1) & 0.666(2) & 0.0352(7) & 0.171(4)\ \ & 40.5(13) \\
7.0\ \ & 1.23(2) & 0.728(2) & 0.0354(7) & 0.205(9)\ \ & 34.1(16) \\ 
6.8\ \ & 1.42(2) & 0.790(2) & 0.0427(6) & 0.257(9)\ \ & 32.7(12) \\
6.25   & 1.75(3) & 0.795(1) & 0.0560(6) & 0.393(13)   & 28.1(10) \\ 
\end{tabular}
\end{center}
\caption{Final fit results from average plaquette simulations.}
\label{PFinal}
\end{table}


\end{document}